\documentclass[conference]{IEEEtran}
\IEEEoverridecommandlockouts
\usepackage{amsmath,amssymb,amsfonts}
\usepackage{algorithmic}
\usepackage{graphicx}
\usepackage{textcomp}
\usepackage{xcolor}
\usepackage{booktabs}
\usepackage{multirow}
\usepackage{cite}
\usepackage{hyperref}

\begin{document}

\title{Nested array design of extended coprime sets for DOA estimation of non-circular signals}

\author{
\IEEEauthorblockN{Dongqi Chen\textsuperscript{1}, Kun Ye\textsuperscript{2}, Chuanxi Xing\textsuperscript{3}, Waqas Khalid\textsuperscript{4}, Huiping Huang\textsuperscript{5}}
\IEEEauthorblockA{
\textsuperscript{1,2,3}School of Electrical and Information Technology, Yunnan Minzu University, Kunming, China\\
Yunnan Key Laboratory of Unmanned Autonomous System, Kunming, China\\
chenliangfaer@163.com, yekun@stu.xmu.edu.cn, xingchuanxi@163.com\\
\textsuperscript{4}Department of Electrical and Electronic Engineering, University of Nottingham Ningbo China, Ningbo, China\\
Waqas.Khalid@nottingham.edu.cn\\
\textsuperscript{5}Department of Electrical Engineering, Chalmers University of Technology, Gothenburg, Sweden\\
huiping@chalmers.se}
}

\maketitle

\begin{abstract}
In recent years, direction of arrival estimation utilizing non-circular signals has become a focal point for scholarly research. To enhance the degrees of freedom (DOF) in receiver arrays specifically for non-circular signal DOA estimation, this study introduces a novel array configuration. This design leverages an extended coprime framework, applying a sliding translation technique to optimize sensor placement. Crucially, this rearranged structure preserves the continuity of the difference co-array (DCA). Furthermore, the sum co-array (SCA) is shifted to merge seamlessly with the DCA, eliminating redundancy and substantially expanding both the virtual aperture array (VAA) and the DOF. Consequently, the proposed array demonstrates superior performance in practical DOA estimation tasks involving non-circular signals. Simulation results and comparative analyses confirm that, relative to traditional Nested Arrays (NA), Extended Sliding Nested Array (ESNA), and other benchmark structures, the proposed array achieves better DOF and VAA, leading to enhanced estimation accuracy in practical scenarios.
\end{abstract}


\section{Introduction}
With the increasingly widespread application of direction of arrival (DOA) in military and commercial fields, it has also attracted more and more attention in array signal processing~\cite{1,2,3,4,5,6,7}. Although the traditional subspace algorithms based on uniform linear arrays can solve the problem, they have significant limitations. Firstly, underdetermined DOA estimation cannot be achieved. That is, for an array composed of \(N\) sensors, at most \(N-1\) source signals can be estimated.

Secondly, the separation of the array sensors is constrained to half a wavelength, which impacts the aperture enlargement of the array and introduces the consequences of mutual coupling effects.

To solve the above problems, sparse arrays have gradually come into the view of researchers, and their research has developed rapidly~\cite{8,9,10,11,12,13}. Minimum redundancy arrays (MRA), coprime array (CPA) and nested arrays (NA) are examples of traditional sparse arrays. The MRA array can break through the sensor spacing of half wavelength and achieve underdetermined DOA estimation. However, no closed-form expression can be given for its sensor position. The NA array not only retains its advantages but also can obtain the closed-form expression of its sensor position, thus receiving much attention in the research of sparse arrays. In addition to the classic NA arrays, a series of variants of NA arrays such as augmented nested arrays (ANAs)~\cite{13}, super nested arrays (SNAs)~\cite{14}, and generalized nested array (GNA)~\cite{15} have emerged, further improving the accuracy of DOA estimation.

Most of the arrays mentioned above are studies for circular signals. In recent years, research on DOA estimation for non-circular signals has also developed rapidly~\cite{16,17,18}. Two relocating sparse nested array (RSNA) structures, RSNA-I and RSNA-II, were presented in~\cite{16}. They can, respectively, reach the degrees of freedom (DOF) of \(4M_1M_2 + 2(M_1 + M_2) + 1\) and \(4M_1M_2+2M_1(2-\bmod(M_1, 2))+2M_2+1-\bmod(M_1, 2)\).

In~\cite{17}, the extended sliding nested array (ESNA) array was designed based on the sliding strategy and can achieve \(2N_1N_2 + 2(N_1 + N_2) -2J + 1 + 4 \lfloor(N_1N_2 + N_2 + 2J)/2\rfloor\) DOF, where \(J = \lceil N_1/2\rceil-1\).

This paper begins by laying out the fundamental principles of the sum-difference co-array (SDCA) concept. Building on this foundation, we’ve developed and put forward a novel nested array employing a sliding strategy, specifically tailored to handle DOA estimation for non-circular signals. Through the analysis and numerical simulation, and compared with NA, ESNA, and RSNA-I arrays. The findings indicate that the array developed in this research has superior accuracy of DOA estimation.

The following are the paper’s primary contributions:
\begin{enumerate}
    \item We design a sliding extended coprime nested array utilizing a sliding methodology. Provide the closed-form expression for the sensor position of the array and the generic formula for DOF.
    \item Through the analysis and numerical simulation, compared with NA, ESNA, and RSNA-I arrays, the novel array configuration detailed in this research paper boasts a better DOF, which translates to more accuracy of DOA estimation outcomes while demonstrating superior performance in challenging low signal-to-noise ratio (SNR) scenarios.
\end{enumerate}

The remaining structure of the paper is as follows: The section II introduces the array receiving model of non-circular signals and the concept of SDCA. The section III presents the array structure and specific details designed in this paper. Section IV is numerical simulation and analysis. The conclusion is given in Section V.

\section{Theoretical Premise}

\subsection{Receiving Signal Model}

Suppose there are \(Q\) signals incident on a receiving array in space. The signal satisfies the assumption of a far-field narrowband signal, with the incident direction being \(\{\theta_1, \theta_2, \cdots, \theta_Q\}\). The receiving array comprises \(M\) sensors, with their positions denoted by \(\mathbf{W}\).

\begin{equation}
\mathbf{W} = \{w_1 d, w_2 d, \cdots, w_M d\}
\label{eq:positions}
\end{equation}

where \(d = \lambda/2\) and \(\lambda\) represents the wavelength of the source signal. The array received signal at time \(t\) can be articulated as:

\begin{equation}
\mathbf{x}(t) = \mathbf{A}\mathbf{s}(t) + \mathbf{e}(t),
\label{eq:received}
\end{equation}

where \(\mathbf{x}(t) = [x_1(t), x_2(t), \cdots, x_M(t)]^T\) is the observed value vector at time \(t\). \(\mathbf{A} = [\mathbf{a}(\theta_1), \mathbf{a}(\theta_2), \cdots, \mathbf{a}(\theta_Q)]\) is the matrix of the array manifold of dimension \(M \times Q\) and

\begin{equation}
\mathbf{a}(\theta_q) = [e^{j\pi w_1 d \sin(\theta_q)}, e^{j\pi w_2 d \sin(\theta_q)}, \cdots, e^{j\pi w_M d \sin(\theta_q)}]^T
\end{equation}

is the \(q\)-th source steering vector. \(\mathbf{s}(t) = [s_1(t), s_2(t), \cdots, s_Q(t)]^T\) is the source signal vector at time \(t\). \(\mathbf{e}(t) = [e_1(t), e_2(t), \cdots, e_M(t)]^T\) is the noise vector at time \(t\), it is assumed to be Gaussian white noise, with variance \(\delta_n^2\) and mean zero.

In this paper, we define the received signal as a non-circular signal and represent it as~\cite{19}:

\begin{equation}
\mathbf{s}_{NC}(t) = \boldsymbol{\Psi} \mathbf{s}_R(t),
\label{eq:noncircular}
\end{equation}

where \(\boldsymbol{\Psi} = \operatorname{diag}\{e^{-j\phi_1}, e^{-j\phi_2}, \cdots, e^{-j\phi_Q}\}\) represents the diagonal matrix of signal delay. \(\phi_q\) represents the phase of the signal with an incident angle of \(\theta_q\). When the number of snapshots is \(t\), the signal vector is represented as \(\mathbf{s}_R(t)\).

Note that the pseudo covariance matrix of the non-circular signals is nonzero~\cite{19}. In order to fully exploit the information conveyed by the non-circular signals, the combination of the original received signal and its conjugate is known as an extended received signal, and it can be written as follows:

\begin{equation}
\mathbf{x}_{NC}(t) = \begin{bmatrix} \mathbf{x}(t) \\ \mathbf{x}^*(t) \end{bmatrix} = \begin{bmatrix} \mathbf{A}\boldsymbol{\Psi} \\ \mathbf{A}^*\boldsymbol{\Psi}^* \end{bmatrix} \mathbf{s}_R(t) + \begin{bmatrix} \mathbf{e}(t) \\ \mathbf{e}^*(t) \end{bmatrix}.
\label{eq:extended}
\end{equation}

The received signal’s extended covariance matrix can be written as:

\begin{equation}
\begin{aligned}
\mathbf{R}_{NC} 
&= E[\mathbf{X}_{NC}\mathbf{X}_{NC}^H] \\
&= \mathbf{A}_{NC} E[\mathbf{S}_R \mathbf{S}_R^H] \mathbf{A}_{NC}^H 
+ \delta_n^2 \mathbf{I}_{2M} \\
&= \mathbf{A}_{NC} \mathbf{R}_{SR} \mathbf{A}_{NC}^H 
+ \delta_n^2 \mathbf{I}_{2M},
\end{aligned}
\label{eq:cov}
\end{equation}
where \(\mathbf{X}_{NC} = [\mathbf{x}_{NC}(1), \mathbf{x}_{NC}(2), \cdots, \mathbf{x}_{NC}(T)]\). \(\mathbf{A}_{NC} = [(\mathbf{A}\boldsymbol{\Psi})^T, (\mathbf{A}\boldsymbol{\Psi})^H]^T\) is defined as the array manifold matrix of the extended signal of dimension \(2M \times Q\). \(\mathbf{R}_{SR} = E[\mathbf{S}_R \mathbf{S}_R^H] = \operatorname{diag}\{\delta_1^2, \delta_2^2, \cdots, \delta_Q^2\}\) is the covariance matrix of the source signal, \(\delta_q^2\) is the energy of the source signal in the \(q\)-th direction.

Vectorization of \(\mathbf{R}_{NC}\) yields the following equation as:

\begin{equation}
\begin{aligned}
\mathbf{z} 
&= \operatorname{vec}(\mathbf{R}_{NC}) 
= (\mathbf{A}_{NC}^* \odot \mathbf{A}_{NC}) \mathbf{s} 
+ \delta_n^2 \operatorname{vec}(\mathbf{I}_{2M}) \\
&= \mathbf{B}_{NC} \mathbf{s} 
+ \delta_n^2 \operatorname{vec}(\mathbf{I}_{2M}),
\end{aligned}
\label{eq:vec}
\end{equation}
where \(\mathbf{s} = [\delta_1^2, \delta_2^2, \cdots, \delta_Q^2]^T\). \(\mathbf{B}_{NC}\) is the array manifold matrix of the SDCA.

However, the phase information in \(\mathbf{B}_{NC}\) is disordered, so a selection matrix \(\mathbf{P}\) is constructed to organize the phase information in \(\mathbf{B}_{NC}\).

\begin{equation}
\mathbf{z}_U = \mathbf{P} \mathbf{z} = \mathbf{B}'_{NC} \mathbf{s} + \delta_n^2 \mathbf{P} \operatorname{vec}(\mathbf{I}_{2M}).
\label{eq:select}
\end{equation}

At this point, \(\mathbf{z}_U\) can be thought of as the received signal of an array that has \(L_U\) sensors. \(L_U\) is the number of array elements of the maximum continuous segment of the SDCA virtual array of the original array. However, its snapshot number is 1. The full-rank covariance matrix is obtained by applying the spatial smoothing procedure for \(\mathbf{z}_U\) based on~\cite{8}. The off-grid sparse Bayesian inference (OGSBI) approach is then used to estimate DOA~\cite{20}.

\subsection{Sum-Difference Co-Array Theoretical Models}

To enable an array with a specific quantity of sensors to achieve better DOF for underdetermined DOA estimation, extending the length of its virtual array with Sum co-array (SCA) and Difference co-array (DCA) is an effective approach.

\textbf{Definition 1:} For an array whose sensor locations set is \(\mathbf{W}\), the sensors locations set \(\mathbf{W}_{SCA}\) of its SCA is defined as follows:

\begin{equation}
\mathbf{W}_{SCA} = \{\pm(W_i + W_j) \mid W_i \in \mathbf{W}, W_j \in \mathbf{W}\}.
\label{eq:sca}
\end{equation}

\textbf{Definition 2:} For an array with sensor locations denoted as \(\mathbf{W}\), the sensor locations set \(\mathbf{W}_{DCA}\) of its DCA is defined as follows:

\begin{equation}
\mathbf{W}_{DCA} = \{W_i - W_j \mid W_i \in \mathbf{W}, W_j \in \mathbf{W}\}.
\label{eq:dca}
\end{equation}

\textbf{Definition 3:} For an array whose sensor locations set is \(\mathbf{W}\), the sensors locations set \(\mathbf{W}_{SDCA}\) of its SDCA is defined as follows:

\begin{equation}
\mathbf{W}_{SDCA} = \mathbf{W}_{SCA} \cup \mathbf{W}_{DCA}.
\label{eq:sdca}
\end{equation}

In subsequent research, DOF specifically refers to the length of the maximum continuous segment of the SDCA in an array. Virtual aperture array (VAA) specifically refers to the SDCA length of the array, that is, \(VAA = \max(\mathbf{W}_{SDCA}) - \min(\mathbf{W}_{SDCA})\).

\section{The Proposed Sliding Extend Coprime Nested Array}

\subsection{Structure of the Sliding Extend Coprime Nested Array}

A sliding extended nested array (SECNA) based on coprime sets will be presented in this section. Firstly, the physical position of the array sensor is given, and then the closed-form expression of the DOF of its SDCA is presented.

The SECNA array consists of three subarrays, and the specific position expressions are as follows:

\begin{equation}
\mathbf{S} = \bigcup_{p=1,2,3} \operatorname{Array}(p),
\label{eq:secna}
\end{equation}

\begin{equation}
\begin{cases}
\operatorname{Array}(1) = N \langle 0 : M-1 \rangle + S \\
\operatorname{Array}(2) = M \langle 0 : N \rangle + S \\
\operatorname{Array}(3) = (N + M) \langle 1 : N + M -1 \rangle + MN + S
\end{cases},
\label{eq:subarrays}
\end{equation}

where \(N\) and \(M\) are a set of coprime numbers. \(S = (M + N)^2 / 2\) is the horizontal sliding length using the sliding strategy. There are \(2(M + N) - 1\) physical sensors in total.

\begin{figure}[htbp]
\centering
\includegraphics[width=\columnwidth]{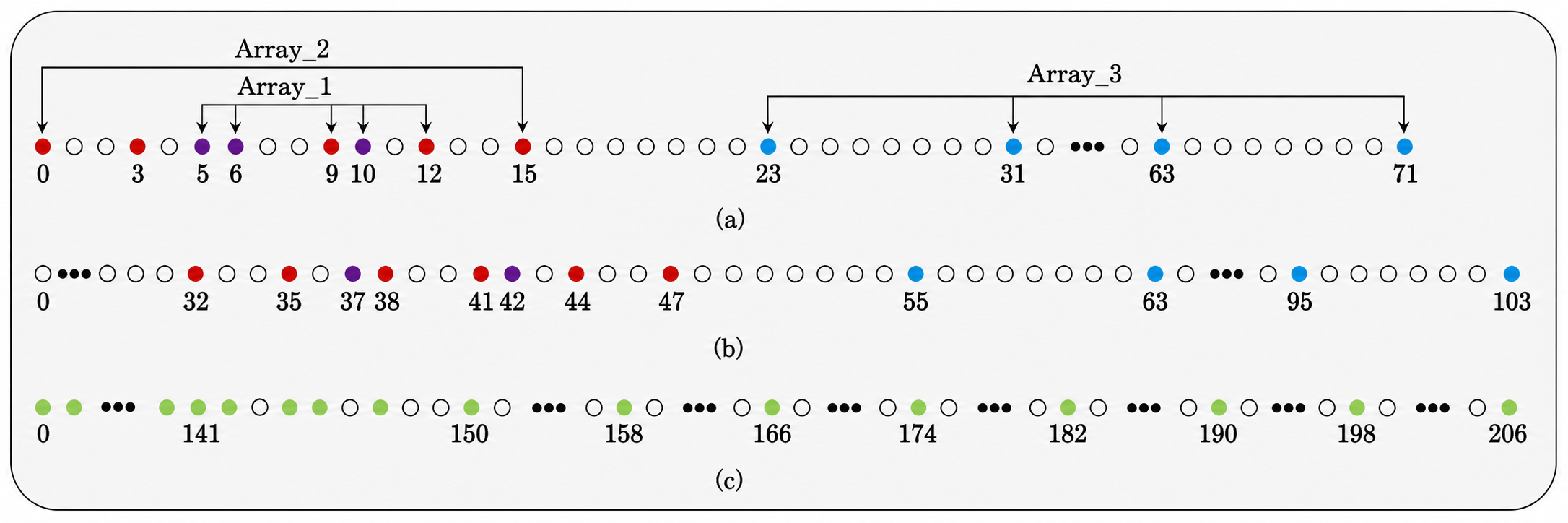}
\caption{(a) The ECNA array structure with \(M = 5\) and \(N = 3\); (b) SECNA: ECNA array structure with sliding strategy; (c) Virtual SDCA on the positive side.}
\label{fig:structure}
\end{figure}

The SECNA structure when the coprime set is \(M = 5\) and \(N = 3\) is depicted in Fig.~\ref{fig:structure}. The three subarrays are respectively: \(\operatorname{Array}(1) = \{32, 35, 38, 41, 44\}\), \(\operatorname{Array}(2) = \{32, 37, 42, 47\}\) and \(\operatorname{Array}(3) = \{55, 63, 71, 79, 87, 95, 103\}\). There are fifteen physical sensors in all. As can be seen from Fig.~\ref{fig:structure}(c), the length of the maximum continuous segment of the positive half-axis of the virtual SDCA array of this array is 142. Therefore, the DOF of the virtual SDCA array composed of 15 physical sensors in the SECNA array can reach 285.

\textbf{Definition 4:} The DOF of SECNA can be calculated in terms of the difference in parity between \(M\) and \(N\):

\begin{equation}
DOF = 4(M + N)^2 + 2MN - 1,
\label{eq:dof}
\end{equation}

\subsection{Comparison of Degrees of Freedom}

This subsection presents a comparison between the DOF of the NA, ESNA, and RSNA arrays and the SECNA array suggested in this paper.

\begin{table}[htbp]
\centering
\caption{DOF Comparison of Different Arrays}
\label{tab:dof}
\begin{tabular}{ccccc}
\toprule
Numbers of sensors & NA & ESNA & RSNA & SECNA \\
\midrule
9  & 61  & 57  & 99  & 111 \\
13 & 113 & 121 & 195 & 219 \\
19 & 221 & 427 & 399 & 441 \\
23 & 313 & 609 & 575 & 645 \\
27 & 421 & 823 & 783 & 873 \\
\bottomrule
\end{tabular}
\end{table}

Table~\ref{tab:dof} presents the specific values of DOF for different arrays, with the number of sensors ranging from 9 to 27. Table~\ref{tab:dof} makes it fairly evident that the DOF of the SDCA of various arrays varies depending on the number of sensors. Among them, the DOF of ESNA and RSNA arrays each have their advantages and disadvantages when the number of elements of array is different. However, the DOF of the SECNA array designed and proposed in this paper are superior to those of NA, ESNA and RSNA when the number of sensors varies. Therefore, the SECNA array is bound to have better working performance when actually conducting DOA estimation.

\section{Numerical Simulation}

In this section, the performance of DOA estimation of different array structures are compared by numerical simulation in terms of root mean square error (RMSE):

\begin{equation}
RMSE = \sqrt{\frac{1}{QK} \sum_{k=1}^{K} \sum_{q=1}^{Q} (\hat{\theta}_{q k} - \theta_q)^2},
\label{eq:rmse}
\end{equation}

where \(K\) is the total number of Monte Carlo simulation trials. \(\hat{\theta}_{q k}\) is the DOA estimate of the signal from direction \(\theta_q\) in the \(k\)-th Monte Carlo simulation trials. In all simulations in this section, \(K\) is set to 200 runs.

The SECNA array created in this study will be compared with the current arrays (including NA, ESNA, and RSNA) in terms of DOA simulation performance. Thirteen physical sensors are included in the array. The estimated number of DOA sources is 31, with an angle ranging from \(-60^\circ\) to \(60^\circ\), following a uniform distribution. Furthermore, the OGSBI method is used in all of the simulations in this research.

\begin{figure}[htbp]
\centering
\includegraphics[width=\columnwidth]{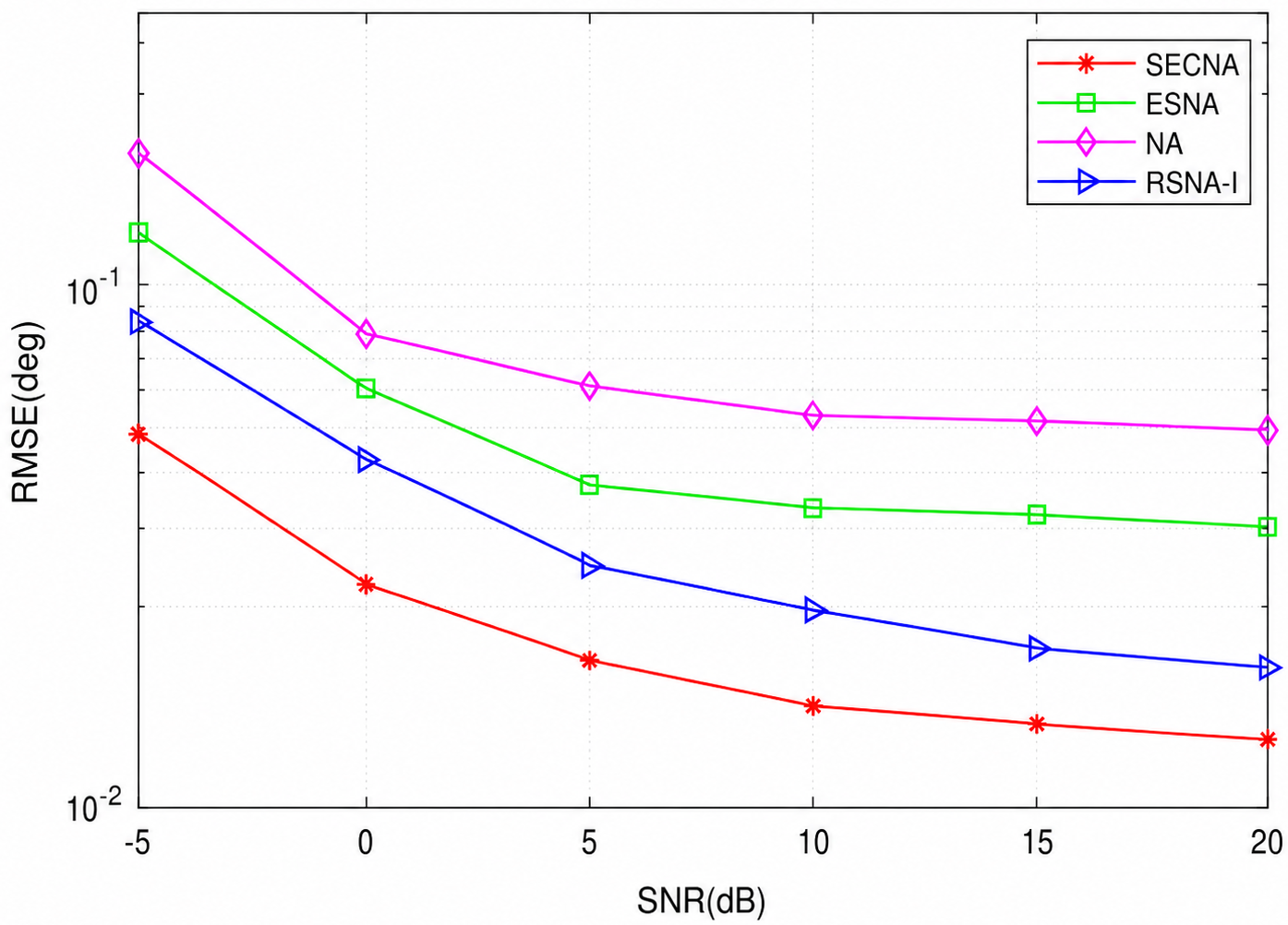}
\caption{RMSE versus SNR}
\label{fig:rmse_snr}
\end{figure}

The comparison of the RMSE of various arrays under various SNRs can be seen in Fig.~\ref{fig:rmse_snr}. The snapshot count is established at 2000. The SNR range is established between \(-5\)dB and \(20\)dB. It is evident that the RMSE of the SECNA array is much lower than that of other arrays under various SNR. RMSE continues to decline as SNR rises. This shows that SECNA has more prominent DOA estimation performance and stronger anti-noise ability.

\begin{figure}[htbp]
\centering
\includegraphics[width=\columnwidth]{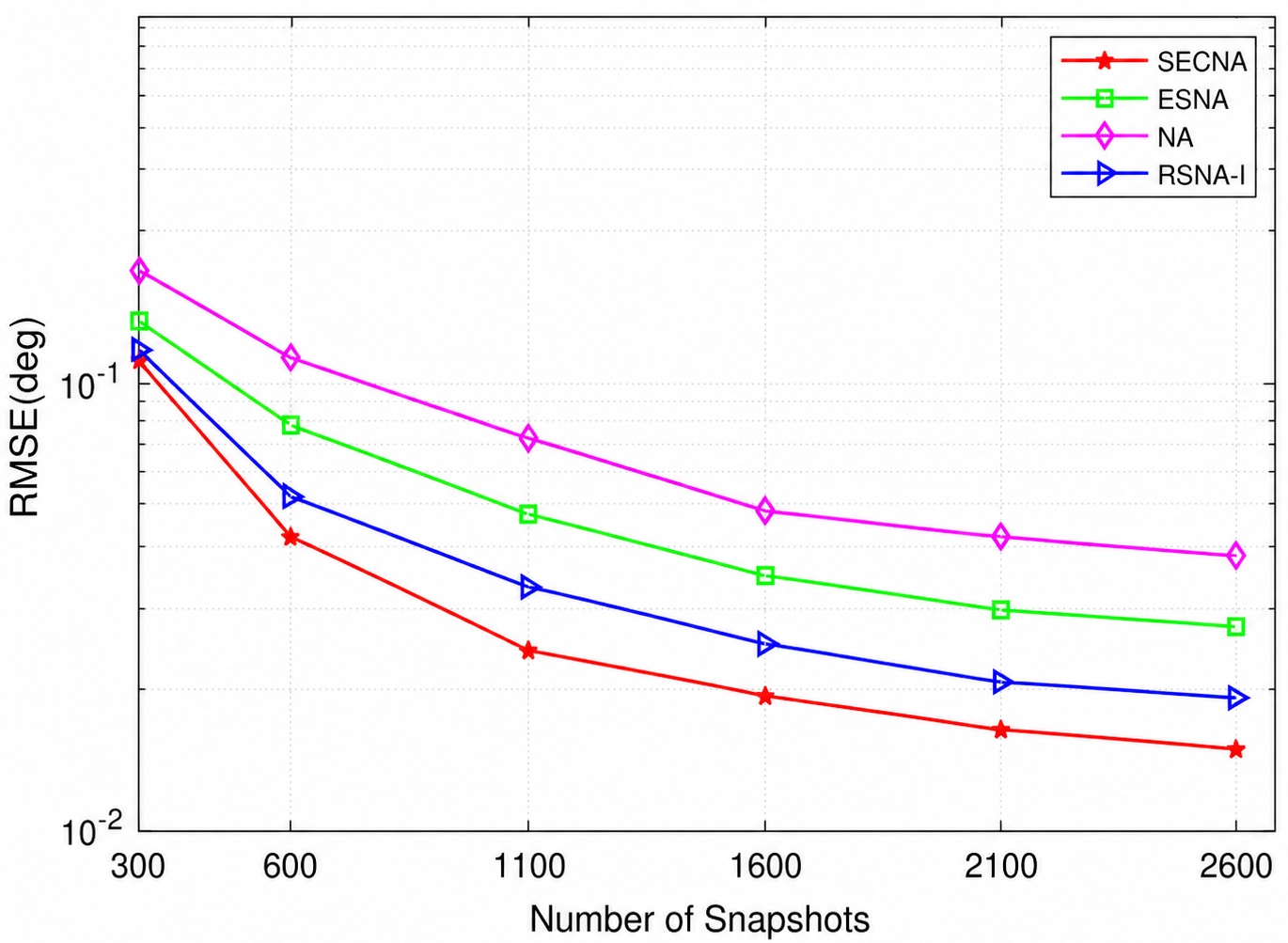}
\caption{RMSE versus snapshots}
\label{fig:rmse_snap}
\end{figure}

The RMSE of various arrays under various snapshots is compared in Fig.~\ref{fig:rmse_snap}. The SNR is set to \(20\)dB. The snapshots range is set to 300 to 2600. It is evident that the SECNA array's RMSE is substantially lower than those of other arrays under various snapshots. As snapshot increases, RMSE further decreases. This shows that SECNA has more prominent DOA estimation performance.

\section{Conclusion}

In this paper, we proposed a novel nested array, better DOF are achieved by using the SDCA in this array, which is based on the sliding strategy. Through numerical analysis and simulation verification, the proposed array has a better DOF and superior estimate performance in DOA estimation where the number of physical sensors is equal to that of NA, ESNA, and RSNA arrays.

\section*{Acknowledgment}

This work was supported by the National Natural Science Foundation of China (No. 61971362).


\begin{thebibliography}{20}

\bibitem{1} C. Zhou, Y. Gu, Z. Shi and Y. D. Zhang, ``Off-Grid Direction-of-Arrival Estimation Using Coprime Array Interpolation,'' in \emph{IEEE Signal Processing Letters}, vol. 25, pp. 1710-1714, Nov. 2018.

\bibitem{2} X. Han, M. Liu, S. Zhang, R. Zheng and J. Lan, ``A Passive DOA Estimation Algorithm of Underwater Multipath Signals via Spatial Time-Frequency Distributions,'' in \emph{IEEE Transactions on Vehicular Technology}, vol. 70, pp. 3439-3455, April 2021.

\bibitem{3} H. Krim and M. Viberg, ``Two decades of array signal processing research: the parametric approach,'' in \emph{IEEE Signal Processing Magazine}, vol. 13, pp. 67-94, July 1996.

\bibitem{4} Z. Zhang, Z. Shi and Y. Gu, ``Ziv-Zakai Bound for DOAs Estimation,'' in \emph{IEEE Transactions on Signal Processing}, vol. 71, pp. 136-149, 2023.

\bibitem{5} G. J. Foschini, G. D. Golden, R. A. Valenzuela and P. W. Wolniansky, ``Simplified processing for high spectral efficiency wireless communication employing multi-element arrays,'' in \emph{IEEE Journal on Selected Areas in Communications}, vol. 17, pp. 1841-1852, Nov. 1999.

\bibitem{6} R. Roy and T. Kailath, ``ESPRIT-estimation of signal parameters via rotational invariance techniques,'' in \emph{IEEE Transactions on Acoustics, Speech, and Signal Processing}, vol. 37, pp. 984-995, July 1989.

\bibitem{7} R. G. Lorenz and S. P. Boyd, ``Robust minimum variance beamforming,'' in \emph{IEEE Transactions on Signal Processing}, vol. 53, pp. 1684-1696, May 2005.

\bibitem{8} P. Pal and P. P. Vaidyanathan, ``Nested Arrays: A Novel Approach to Array Processing With Enhanced Degrees of Freedom,'' in \emph{IEEE Transactions on Signal Processing}, vol. 58, pp. 4167-4181, Aug. 2010.

\bibitem{9} P. P. Vaidyanathan and P. Pal, ``Sparse Sensing With Co-Prime Samplers and Arrays,'' in \emph{IEEE Transactions on Signal Processing}, vol. 59, pp. 573-586, Feb. 2011.

\bibitem{10} A. Raza, W. Liu and Q. Shen, ``Thinned Coprime Array for Second-Order Difference Co-Array Generation With Reduced Mutual Coupling,'' in \emph{IEEE Transactions on Signal Processing}, vol. 67, no. 8, pp. 2052-2065, April 2019.

\bibitem{11} W. Wang, S. Ren, Z. Chen, ``Unified coprime array with multi-period subarrays for direction-of-arrival estimation,'' in \emph{Digital Signal Processing}, Vol. 74, pp. 30-42, March 2018.

\bibitem{12} X. Wang and X. Wang, ``Hole Identification and Filling in k-Times Extended Co-Prime Arrays for Highly Efficient DOA Estimation,'' in \emph{IEEE Transactions on Signal Processing}, vol. 67, pp. 2693-2706, May 2019.

\bibitem{13} J. Liu, Y. Zhang, Y. Lu, S. Ren and S. Cao, ``Augmented Nested Arrays With Enhanced DOF and Reduced Mutual Coupling,'' in \emph{IEEE Transactions on Signal Processing}, vol. 65, pp. 5549-5563, Nov. 2017.

\bibitem{14} C.-L. Liu and P. P. Vaidyanathan, ``Super Nested Arrays: Linear Sparse Arrays With Reduced Mutual Coupling—Part I: Fundamentals,'' in \emph{IEEE Transactions on Signal Processing}, pp. 3997-4012, Aug. 2016.

\bibitem{15} J. Shi, G. Hu, X. Zhang and H. Zhou, ``Generalized Nested Array: Optimization for Degrees of Freedom and Mutual Coupling,'' in \emph{IEEE Communications Letters}, vol. 22, pp. 1208-1211, June 2018.

\bibitem{16} L. Zhou, Z. Feng, K. Ye, ``Design of relocating sparse nested arrays for DOA estimation of non-circular signals,'' in \emph{AEU-International Journal of Electronics and Communications}, vol. 173, pp. 154976, January 2024.

\bibitem{17} X. Zhang, X. Lai, W. Zheng and Y. Wang, ``Sparse Array Design for DOA Estimation of Non-Circular Signals: Reduced Co-Array Redundancy and Increased DOF,'' in \emph{IEEE Sensors Journal}, vol. 21, pp. 27928-27937, Dec. 2021.

\bibitem{18} K. Ye, L. Zhou, Z. Chen, ``DOA estimation based on a novel shifted coprime array structure,'' in \emph{AEU-International Journal of Electronics and Communications}, vol. 179, pp. 155308, May 2024.

\bibitem{19} H. Abeida and J.-P. Delmas, ``MUSIC-like estimation of direction of arrival for noncircular sources,'' in \emph{IEEE Transactions on Signal Processing}, vol. 54, pp. 2678-2690, July 2006.

\bibitem{20} Z. Yang, L. Xie and C. Zhang, ``Off-Grid Direction of Arrival Estimation Using Sparse Bayesian Inference,'' in \emph{IEEE Transactions on Signal Processing}, vol. 61, pp. 38-43, Jan. 2013.

\end{thebibliography}
\end{document}